\documentclass[12pt,preprint]{aastex}

\shorttitle{Pierre Auger UHE neutrinos from point-like sources}

\shortauthors{The Pierre Auger Collaboration}

\begin{document}

\title{ 
SEARCH FOR POINT-LIKE SOURCES OF ULTRA-HIGH ENERGY NEUTRINOS AT THE PIERRE AUGER OBSERVATORY AND IMPROVED LIMIT ON THE DIFFUSE FLUX OF TAU NEUTRINOS}

\author{{\bf THE PIERRE AUGER COLLABORATION}$^\dag$, \\
P.~Abreu$^{63}$, 
M.~Aglietta$^{51}$, 
M.~Ahlers$^{94}$, 
E.J.~Ahn$^{81}$, 
I.F.M.~Albuquerque$^{15}$, 
D.~Allard$^{29}$, 
I.~Allekotte$^{1}$, 
J.~Allen$^{85}$, 
P.~Allison$^{87}$, 
A.~Almela$^{11,\: 7}$
% ...
}

\affil{$^\dag$Observatorio Pierre Auger, Av. San Mart\'{\i}n Norte 304,
5613 Malarg\"ue, Argentina \\
$^{1}$ Centro At\'{o}mico Bariloche and Instituto Balseiro (CNEA-UNCuyo-CONICET), San 
Carlos de Bariloche, 
Argentina \\
% ...
$^{7}$ Instituto de Tecnolog\'{\i}as en Detecci\'{o}n y Astropart\'{\i}culas (CNEA, CONICET, UNSAM), 
Buenos Aires, 
Argentina \\
% ...
$^{11}$ Universidad Tecnol\'{o}gica Nacional - Facultad Regional Buenos Aires, Buenos Aires,
Argentina \\
% ...
$^{15}$ Universidade de S\~{a}o Paulo, Instituto de F\'{\i}sica, S\~{a}o Paulo, SP, 
Brazil \\
% ...
$^{29}$ Laboratoire AstroParticule et Cosmologie (APC), Universit\'{e} Paris 7, CNRS-IN2P3, 
Paris, 
France \\
% ...
$^{51}$ Istituto di Fisica dello Spazio Interplanetario (INAF), Universit\`{a} di Torino and 
Sezione INFN, Torino, 
Italy \\
% ...
$^{63}$ LIP and Instituto Superior T\'{e}cnico, Technical University of Lisbon, 
Portugal \\
% ...
$^{81}$ Fermilab, Batavia, IL, 
USA \\
% ...
$^{85}$ New York University, New York, NY, 
USA \\
% ...
$^{87}$ Ohio State University, Columbus, OH, 
USA \\
% ...
$^{94}$ University of Wisconsin, Madison, WI, 
USA \\
% ...
(97) Deceased \\
(98) now at University of Maryland, USA \\
(99) now at Institut de F\'{\i}sica d'Altes Energies, Bellaterra, Spain \\
(100) now at Universit\'{e} de Lausanne, Switzerland \\
(101) now at Konan University, Kobe, Japan \\
(102) now at NYU Abu Dhabi \\
(103) now at the Universidad Autonoma de Chiapas on leave of absence from Cinvestav \\
}



\begin{abstract}

The Surface Detector array of the Pierre Auger Observatory can detect neutrinos with energy $E_\nu$ between $10^{17}$ eV and $10^{20}$ eV
from point-like sources across the sky south of $+55^\circ$ and north of $-65^\circ$ declinations. A search has been performed for highly inclined extensive air showers produced by the interaction of neutrinos of all flavours in the atmosphere (downward-going neutrinos), and by the decay of tau leptons  originating from tau neutrinos interactions in the Earth's crust (Earth-skimming neutrinos). No candidate neutrinos have been found in data up to 2010 May 31.
This corresponds to an equivalent exposure of $\sim$3.5 years of a full surface detector array for the Earth-skimming channel and $\sim$2 years for the downward-going channel.
An improved upper limit on the diffuse flux of tau neutrinos has been derived. Upper limits on the neutrino flux from point-like sources have been derived as a function of the source declination.
Assuming a  differential neutrino flux $k_{PS} \cdot E^{-2}_\nu$ from a point-like source,  $90\%$ C.L. upper limits for $k_{PS}$ at the level of $\approx 5 \times 10^{-7} $ and $2.5 \times 10^{-6} \hspace{0.1cm}~{\rm GeV~cm^{-2}~s^{-1}}$ have been obtained over a broad range of declinations from the  searches of Earth-skimming and downward-going neutrinos, respectively.
\end{abstract}

\section{ INTRODUCTION }
\label{sec:Intro}

The nature and production mechanisms of ultra-high energy cosmic rays (UHECRs), with energies above $10^{18}$ eV,
are still unknown~\cite{Nagano00,Bhatta00,Halzen02}. The observation of UHECRs makes an associated flux
of ultra-high energy cosmic neutrinos (UHE$\nu$s)~\cite{Becker08} very likely.
All models of UHECR production predict neutrinos as a result of the decay
of charged pions generated in interactions of cosmic rays within the sources
themselves (``astrophysical'' neutrinos), and/or
in their propagation through background radiation fields
(``cosmogenic'' neutrinos)~\cite{Berez69,Stecker73}.
In fact, charged pions, which are photoproduced by UHECR protons interacting with the Cosmic Microwave Background radiation, decay into UHE$\nu$s.  However, the predicted flux has large uncertainties, since it depends on the UHECR spectrum and on the spatial distribution and cosmological evolution of the sources~\cite{Becker08,Ahlers10,Kotera10}.
If UHECRs are heavy nuclei, the UHE$\nu$ yield is strongly suppressed~\cite{Ave05}.

The observation of UHE neutrinos would open a new window to the Universe.
Neutrinos travel unaffected by magnetic fields
and can give information on astrophysical regions that are otherwise
hidden from observation by large amounts of matter.
The discovery of astrophysical neutrino sources
would shed light on the long-standing question of the
origin of cosmic rays, and clarify the production mechanism
 of the GeV-TeV gamma-rays observed on Earth~\cite{Gaisser95, Alvarez02}.

The Pierre Auger Observatory~\cite{Abraham04} -- located in the province
of Mendoza, Argentina, at a mean altitude of 1400~m above sea level ($\sim$875 g cm$^{-2}$) --
was designed to measure extensive air showers (EAS) induced by
UHECRs.
The Fluorescence Detector (FD) \cite{Abraham10a} comprises a set of imaging telescopes to
measure the light emitted by excited atmospheric nitrogen molecules
as the EAS develops. A Surface Detector (SD)~\cite{Allekotte08},
measures EAS particles at ground with an array of water-Cherenkov
detectors (``stations''). Each SD station contains 12 tonnes of
water viewed by three 9'' photomultipliers (PMTs). Arranged on a
triangular grid with 1.5 km spacing, 1660 SD stations are deployed
over an area of $\sim 3000~{\rm km^2}$, overlooked by 27
fluorescence telescopes.

Although the primary goal of the SD is to detect UHECRs, it can also
identify ultra-high energy neutrinos. Neutrinos of all flavours can
interact at any atmospheric depth through charged or neutral
currents and induce a ``downward-going'' ({\it DG}) shower. In
addition, tau neutrinos can undergo charged current interactions in
the Earth crust and produce a tau lepton which, after emerging from
the Earth surface and decaying in the atmosphere, will induce an
``Earth-skimming'' ({\it ES}) upward-going shower. Even if  tau
neutrinos are not expected to be produced at the astrophysical
source, approximately equal fluxes for each neutrino flavour should
reach the Earth as a result of neutrino oscillations over
cosmological distances. Neutrino candidate events must be identified
against the overwhelming background of showers initiated by standard
UHECRs (protons or nuclei) and, in a much smaller proportion,
photons~\cite{Abraham10b}. Highly inclined  (zenith angle  $\theta >
75^\circ$) {\it ES}  and {\it DG} neutrino-induced showers will
present a significant electromagnetic component at the ground
(``young'' showers), producing signals spread over hundreds of
nanoseconds in several of the triggered SD stations. Inclined
showers initiated by standard UHECRs are, by contrast, dominated by
muons at ground level (``old'' showers), with signals typically
spread over only tens of nanoseconds.
Thanks to the fast sampling (25 ns) of the SD digital electronics,
several observables sensitive to the signal time structure can be
used to discriminate between young and old showers,  allowing for
detection of UHE$\nu$s. Candidates for UHE$\nu$s are searched for in
inclined showers in the ranges $75^\circ<\theta<90^\circ$ and
$90^\circ<\theta<96^\circ$ for the {\it DG}  and {\it ES}  analysis,
respectively.

\section{ LIMITS ON THE DIFFUSE FLUX OF UHE TAU NEUTRINOS }
\label{sec:DiffLimits}

An upper limit on the diffuse flux of tau neutrinos from the search  of Earth-skimming events in data through 2008 April 30 ($\sim$2 years of exposure with a full SD array) was reported in \cite{Abraham09}. Here, the search is extended to data until 2010 May 31 ($\sim$3.5 years of exposure with a full SD array), and an improved limit is obtained. A preliminary report of this result was presented in~\cite{Yann11}.

\begin{figure}
\begin{center}
\includegraphics[angle=0.,scale=.70]{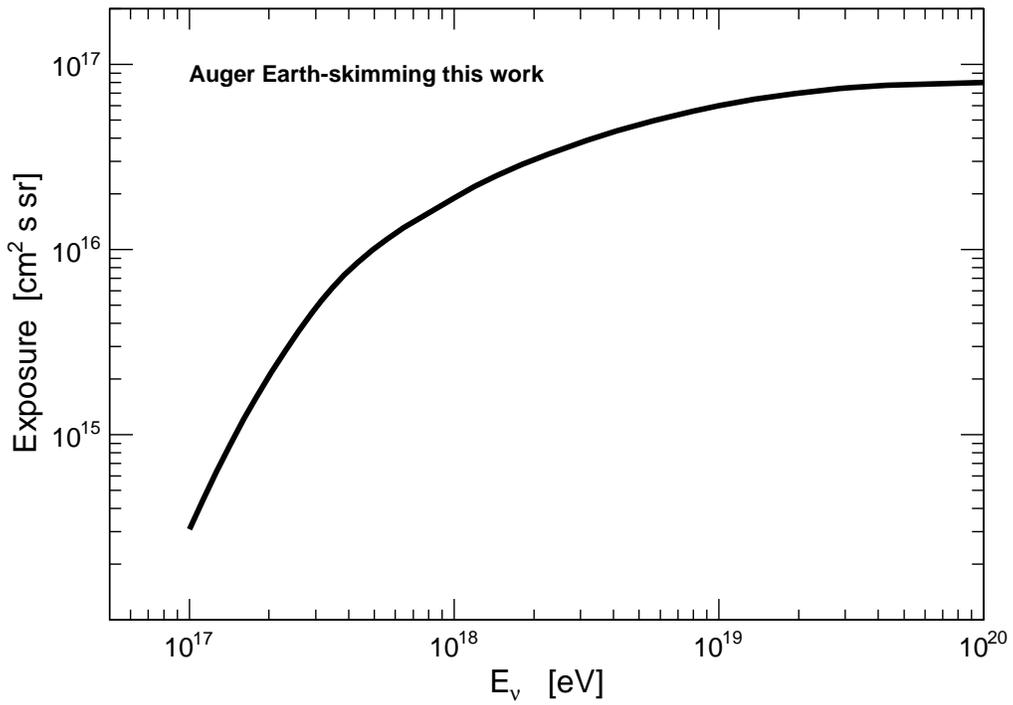}
\caption{
Exposure of the Surface Detector of the Pierre Auger Observatory for Earth-skimming neutrino initiated showers as a function of the neutrino energy, for
data collected between 2004 January 1 and 2010 May 31.
\label{fig:exposurediff}}
\end{center}
\end{figure}

\begin{figure}
\begin{center}
\includegraphics[angle=0.,scale=.80]{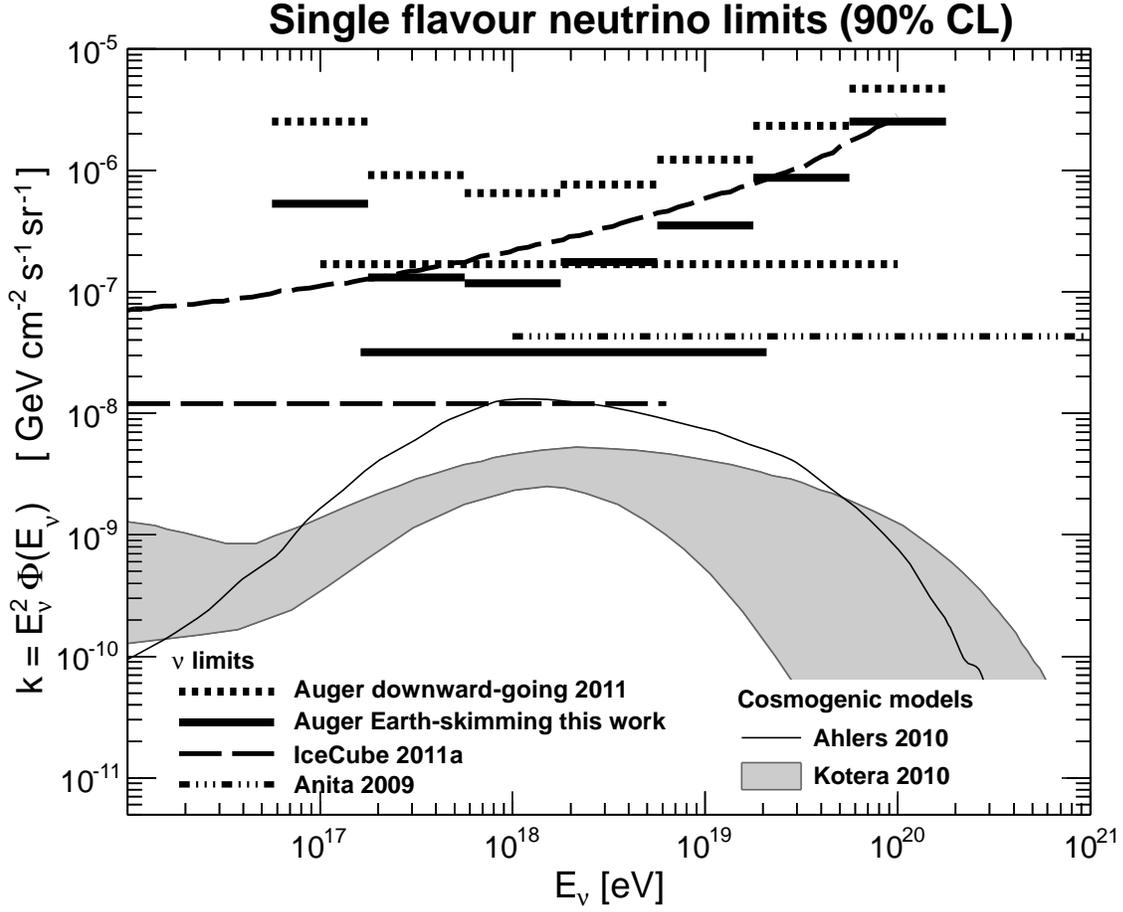}
\caption{
Differential and integrated upper limits at 90$\%$ C.L.
on the single flavour E$^{-2}_\nu$ neutrino flux from the search for downward-going and Earth-skimming neutrinos at the Pierre Auger Observatory.
Integrated upper limits are indicated by horizontal lines, with the corresponding differential limits being represented by segments
of width $0.5$ in $\log_{10}E_\nu$.
Limits from the IceCube Neutrino Observatory~\cite{Abbasi11a}
and from the ANITA experiment~\cite{Gorham10}
are also shown after proper rescaling to account for single
flavour neutrino flux and different energy binning.
Predictions for cosmogenic neutrinos under different assumptions~\cite{Ahlers10, Kotera10}
are also shown, although predictions almost one order of magnitude lower or higher exist.
\label{fig:DiffLimit}}
\end{center}
\end{figure}

Details of the neutrino selection procedure, of the calculation of the detector exposure for {\it ES} showers, and of sources of systematic uncertainties are given in~\cite{Abraham09}.
The neutrino selection criteria were optimized with an early data set collected  between 2004 November 1 and 2004 December 31. By using data rather than Monte Carlo simulations, all possible detector effects and shower-to-shower fluctuations, which constitute the main background to UHE$\nu$s and may not be well reproduced by simulations, are taken into account. The neutrino selection established with the training sample was then applied to  a ``blind search sample'' of data collected between 2004 January 1 and 2010 May 31 (excluding November and December 2004). The blind search sample is equivalent to $\sim 3.5$ years of data collected by a fully efficient SD array, {\it i.e.} with all stations working continuously.
The time evolution of the SD array, which was growing during the construction phase, as well as the dead times of individual stations, were accounted for in this calculation. The integrated exposure for detection of  {\it ES} tau neutrinos as a function of energy is shown in Figure~\ref{fig:exposurediff}. No neutrino candidates were found in the blind search. Assuming a differential spectrum  $\Phi(E_\nu) = dN_\nu/dE_\nu = k\cdot E_\nu^{-2}$ for the  diffuse flux of UHE$\nu$s and zero background \cite{Abraham09, Abreu11a}, a 90\% C.L. upper limit on the integrated flux of tau neutrinos is derived:

\begin{equation}
k < 3.2 \times 10^{-8}~\hspace{0.3cm}~{\rm GeV~cm^{-2}~s^{-1}~sr^{-1}}.
\label{eq:kLimit}
\end{equation}
Systematic uncertainties in the exposure were taken into account in
the upper limit by using a semi-Bayesian extension~\cite{Conrad03}
of the Feldman-Cousins approach~\cite{Feldman98}. The limit, shown
as a horizontal line in Figure~\ref{fig:DiffLimit}, is valid in the
energy range $1.6\times10^{17}~{\rm eV} \leq E_\nu \leq
2.0\times10^{19}~{\rm eV}$, where $\approx 90\%$ of neutrino events
would be detected for a $E_\nu^{-2}$ flux . Also shown is the $90\%$
C.L. upper limit  in differential form, where the limit was
calculated independently in each energy bin of width $0.5$ in
$\log_{10}E_\nu$. The integrated and differential limits from the
search for downward-going neutrinos~\cite{Abreu11a} at the Pierre
Auger Observatory, based on a ``blind search sample'' of data
collected from 2007 November 1 until 2010 May 31 (equivalent to
$\sim 2.0$ years of exposure with the full SD array), are also shown
in Figure~\ref{fig:DiffLimit}, together with limits from the IceCube
Neutrino Observatory~\cite{Abbasi11a} and the ANITA
experiment~\cite{Gorham10}.
The shaded area in Figure~\ref{fig:DiffLimit} brackets the cosmogenic neutrinos fluxes predicted under a wide range of assumptions for the cosmological evolution of the sources, for the transition between the galactic and extragalactic component of cosmic rays, and for the UHECR composition~\cite{Kotera10}. The corresponding number of cosmogenic neutrino events expected in the blind search sample ranges between 0.1 and 0.3, approximately.
For the diffuse flux of cosmogenic neutrinos predicted in~\cite{Ahlers10}, $0.6$ neutrino events are expected at the Pierre Auger Observatory with the integrated exposure of the present analysis, to be compared with $0.43$  events expected in the $333.5$ days of live-time of the IceCube-40 neutrino telescope~\cite{Abbasi11a}.
The current bound to a cosmogenic neutrino flux with energy dependence as in~\cite{Ahlers10} and shown in Figure~\ref{fig:DiffLimit}
is 4 times larger than the predicted value. With the current selection criteria the exposure accumulated in $\sim$10 more years
with the Pierre Auger Observatory may exclude this cosmogenic neutrino flux at 90\% C.L..
Notice that the maximum sensitivity of the Pierre Auger Observatory, obtained for $E_\nu \sim 10^{18}$ eV, matches well the peak of the expected neutrino flux.

\section{ SENSITIVITY TO POINT-LIKE SOURCES }
\label{sec:Sensitivity}

\begin{figure}
\begin{center}
\includegraphics[angle=0.,scale=.80]{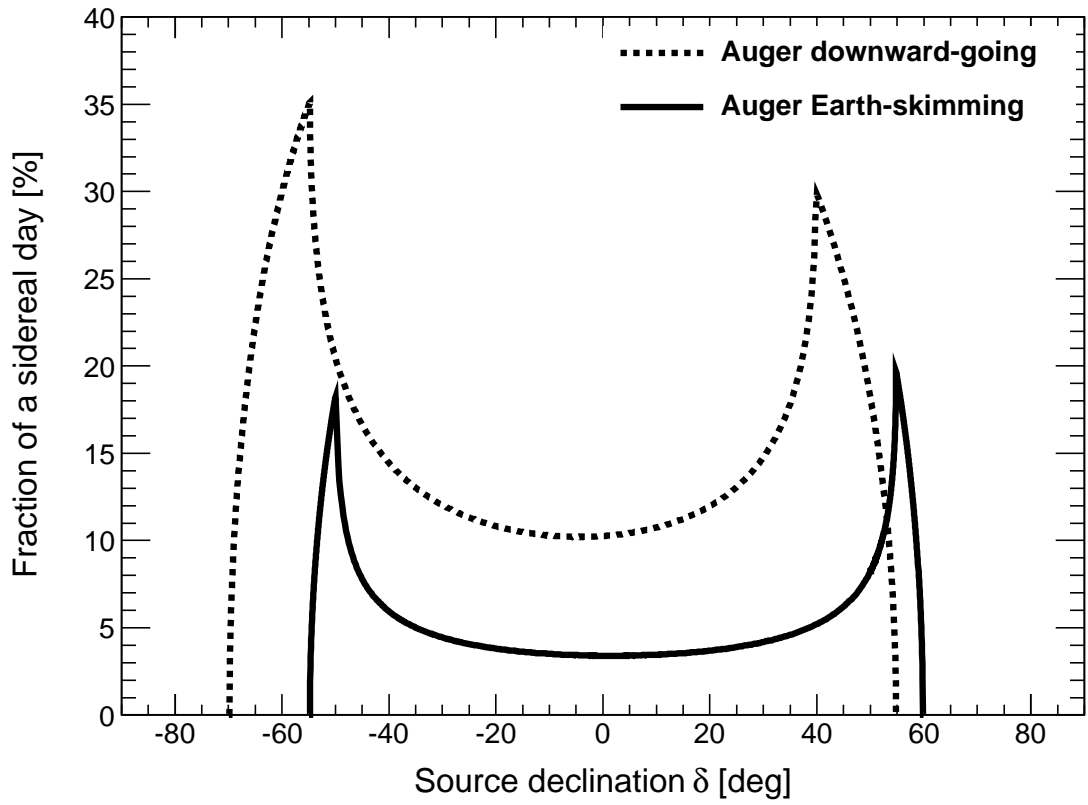}
\caption{Fraction of a sidereal day having a point-like source at declination
  $\delta$ detectable by the Pierre Auger Observatory
with the Earth-skimming and downward-going neutrino selection.
\label{fig:PS-FractimeV}}
\end{center}
\end{figure}

The neutrino search at the Pierre Auger Observatory is limited to
highly inclined showers, with zenith angles between $90^\circ$ and
$96^\circ$ in the Earth-skimming analysis, and between $75^\circ$
and $90^\circ$ in the downward-going analysis. Thus, at each
instant, neutrinos can be detected only from a specific portion of
the sky corresponding to these zenith angle ranges. A point-like
source of declination $\delta$ and right ascension $\alpha$
(equatorial coordinates) is seen at our latitude ($\lambda =
-35.2^\circ$), at a given sidereal time $t$, with a zenith angle
$\theta(t)$ given by:
\begin{equation}
\cos\theta(t) = \sin\lambda\,\sin\delta +
\cos\lambda\,\cos\delta\,\sin(2 \pi t/T - \alpha) \,\, ,
\label{eq:costheta-t}
\end{equation}
where $T$ is the duration of one sidereal day. From
equation~\ref{eq:costheta-t}, the fraction of a sidereal day during
which a source is detectable (i.e., within the zenith angle ranges
for the {\it ES} and {\it DG} analyses) is shown in
Figure~\ref{fig:PS-FractimeV}; it depends only on the source
declination. The SD of the Pierre Auger Observatory is sensitive to
point-like sources of neutrinos over a broad declination range
spanning north of $\delta\sim -65^\circ$ and south of $\delta \sim
55^\circ$. The regions of the sky close to the Northern
($\delta=90^\circ$) and Southern ($\delta=-90^\circ$) Terrestrial
Poles are not accessible by this analysis. As an example, Centaurus
A ($\delta\sim-43^\circ$) is observed $\sim 7\%$ ($\sim 15\%$) of
one sidereal day in the range of zenith angles corresponding to the
{\it ES} ({\it DG}) search. The peaks in
Figure~\ref{fig:PS-FractimeV} are a consequence of the relatively
smaller rate of variation of zenith angle with time for directions
near the edges of the range accessible to this analysis.

The exposure of the SD as a function of the neutrino energy and of
the source position in the sky, ${\cal E}(E_\nu, \delta)$, is
evaluated by folding the SD aperture with the neutrino interaction
probability and the selection efficiency for each neutrino channel.
The procedure is identical to that used for the calculation of the
exposure for a diffuse flux of
UHE$\nu$s~(\cite{Abraham09,Abreu11a,Yann11}), with the exception of
the solid angle integration over the sky. The integration over the
blind search time period takes into account the growth of the SD
array during its construction phase and dead times of individual
stations. For example,  the exposure for the  {\it DG} analysis is
given by:

\begin{equation}\label{eq:exposure}
{\cal E}(E_\nu,\delta)= \frac{1}{m}\sum_i \left[\omega_i ~ \sigma_i(E_\nu)
 \int\int\!\int\! \cos\theta(t) ~ \varepsilon_i(\vec{r},E_\nu,\theta(t),D,t)~{\rm d}A~{\rm d}D~{\rm d}t \right]
\end{equation}

%
where the integration is performed over the area $A$ of the SD,
the interaction depth $D$ of the neutrino, and the search period.
In equation~\ref{eq:exposure},  $m$ is the mass of a nucleon, $\sigma_i(E_\nu)$ is the neutrino-nucleon
cross-section~\cite{Sarkar08}, and $\varepsilon_i$ is the neutrino selection efficiency, with the sum running over the 3 neutrino flavours ($\omega_i=1$,
corresponding to a 1:1:1 flavour ratio) and over the neutrino charged and neutral current interactions.
The dependence of $\varepsilon$ on several parameters (the point of impact at ground of the shower core, $\vec r$, the neutrino interaction depth, its energy and zenith angle, and
time) is also explicitly included in equation~\ref{eq:exposure}.
The dependence of the exposure on the source declination comes through $\theta(t)$ as obtained from equation~\ref{eq:costheta-t}. When integrating over time, only those periods when the source is within the zenith angle range of the neutrino selection are considered.
The exposure for ES neutrinos is derived analogously to equation~\ref{eq:exposure}.

Changes in the detector configuration during data taking,
due to the dead times of the SD stations, and to the increase of the array size
during the construction phase, may introduce a dependence of the exposure
on the right ascension.
In particular, fluctuations in the number of stations cause a small
diurnal variation, but this effect is only apparent in solar time.
When averaged over a large number of sidereal days, as in this
analysis, the modulation in right ascension caused by this effect is
less than 1\% ~\cite{Abreu11b}. For this reason, the dependence of
the exposure on $\alpha$ has been neglected in the evaluation of the
upper limits.

Due to the finite resolution of the SD on the reconstruction of the
variables used in the selection of neutrino-induced showers, events close
to the edges of the zenith angle range for the neutrino selection may be
wrongly rejected (or wrongly accepted). In the exposure as given in
equation~\ref{eq:exposure} we account for this effect by evaluating the selection
efficiency $\varepsilon$  through Monte Carlo simulations.

Several other sources of systematic uncertainties on the exposure have been investigated (\cite{Abraham09, Abreu11a}).
For the  {\it DG} analysis, the major contributions in terms of deviation from a reference exposure
come from the knowledge of neutrino-induced shower simulations ($+9\%, -33\%$),
of the neutrino cross-section ($\pm 7\%$), and of the topography ($\pm 6\%$).
Only uncertainties compatible with the conventional NLO DGLAP formalism
of $\nu$ cross-section calculation -- see \cite{Sarkar08}
for details -- have been considered. We have not accounted for
gluon saturation models that would give rise to considerable smaller $\nu$ cross-sections
(as small as a factor $\sim 2$ at $10^{18}$ eV \cite{Henley06, Armesto08}),
and hence to a larger systematic uncertainty than the one quoted here.
For the {\it ES} analysis, the systematic uncertainties are dominated by
the tau energy losses ($+25\%, -10\%$), the shower simulations ($+20\%, -5\%$) and
the topography ($+18\%, 0\%$).

\section{ LIMITS ON THE FLUX OF UHE NEUTRINOS FROM POINT-LIKE SOURCES }
\label{sec:Results}

The expected number of neutrino events in an energy range $\left[
E_{\rm min},E_{\rm max} \right] $ from a point-like source located
at a declination $\delta$ is given by:

\begin{equation}\label{eq:Nevents}
N^{\rm point~source}_{\rm expected}(\delta)=
\int_{E_{\rm min}}^{E_{\rm max}} F(E_{\nu})\,{\cal E}(E_{\nu},\delta)\,{\rm d}E_{\nu} \,\, ,
\end{equation}
where $F(E_{\nu})$ is the flux of UHE$\nu$s from the source. No
candidate events were selected using the  {\it ES} and {\it DG}
analyses. Under the conservative assumption of zero background, a
90\% C.L. upper limit on the neutrino flux from point-like sources
is derived.  To set the upper limit, a differential flux $F(E_\nu )
= k_{PS}(\delta) \cdot E^{-2}_\nu$ was assumed, as well as a 1:1:1
neutrino flavour ratio. Systematic uncertainties on the exposure
were calculated using the semi-Bayesian approach described above in
section 2.

\begin{figure}
\begin{center}
\includegraphics[angle=0.,scale=.80]{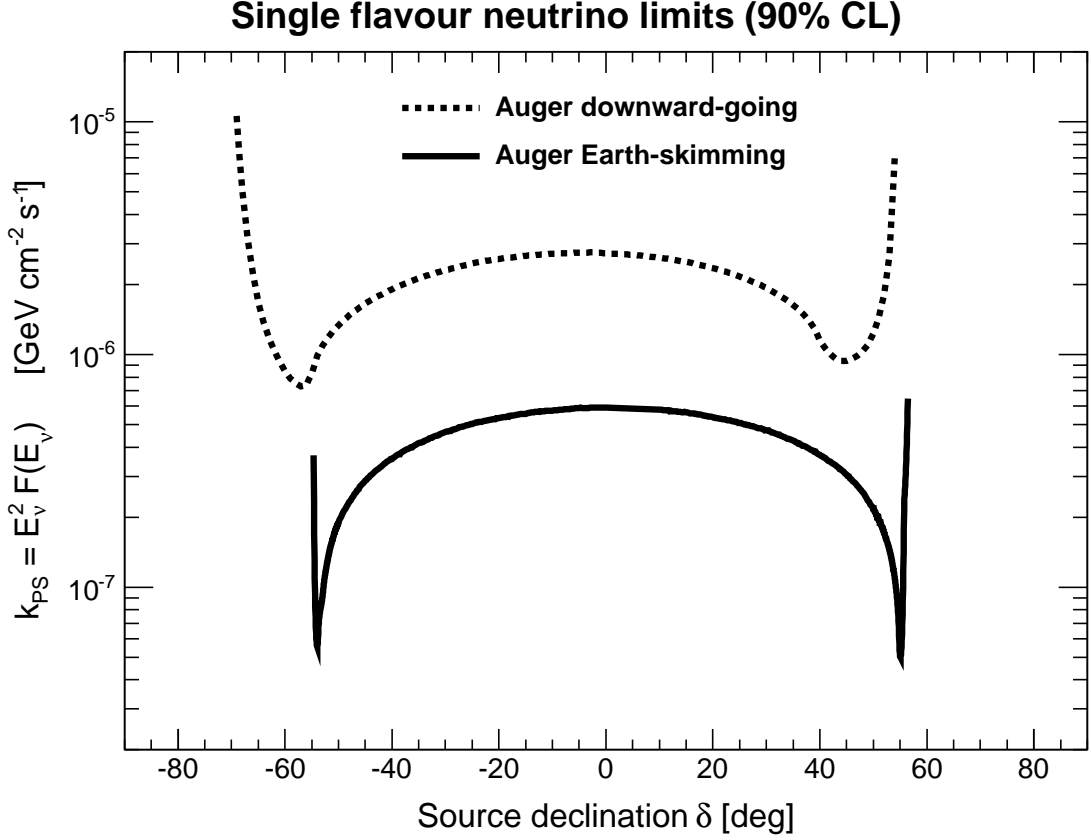}
\caption{Upper limits at 90$\%$ C.L.  on a single flavour $E^{-2}_\nu$ flux from a specific point-like source
as a function of the source declination. The bounds from the Earth-skimming and downward-going neutrino analyses hold for a neutrino energy range  $10^{17} - 10^{20}$ eV (see text for details).
\label{fig:k90_limit_dec}}
\end{center}
\end{figure}

In Figure~\ref{fig:k90_limit_dec}, the $90\%$ C.L. upper limits on $k_{PS}$ derived from the {\it ES} and {\it DG} analyses are shown
as a function of source declination.
Limits for $k_{PS}$ at the level of $\approx 5 \times 10^{-7} $ and $2.5 \times 10^{-6} \hspace{0.1cm}~{\rm GeV~cm^{-2}~s^{-1}}$ were obtained over a broad range of declinations from the  searches of Earth-skimming and downward-going neutrinos, respectively.

The shape of the declination-dependent upper limits is largely determined by the fraction of time a source is within the field of view of the {\it ES} or {\it DG} analyses (cf. Figure~\ref{fig:PS-FractimeV}), and, to a lesser extent, by the zenith angle dependence of the exposure.

The upper limits are derived for neutrinos in the energy range
$1.6\times10^{17}~{\rm eV} - 2.0\times10^{19}~{\rm eV}$ for the
Earth-skimming analysis, and in the energy range
$1\times10^{17}~{\rm eV} - 1\times10^{20}~{\rm eV}$ for the downward-going
analysis, with a negligible dependence of these energy intervals on the source
declination. These are the best limits around 1~EeV.

The IceCube Neutrino Observatory and the Antares Neutrino Telescope have also searched for UHE$\nu$s
from point-like sources (\cite{Abbasi11b} and~\cite{Adrian11}, respectively).
The bounds obtained by these two experiments apply to energies below the Auger energy range.

\begin{figure}
\begin{center}
\includegraphics[angle=0.,scale=0.8]{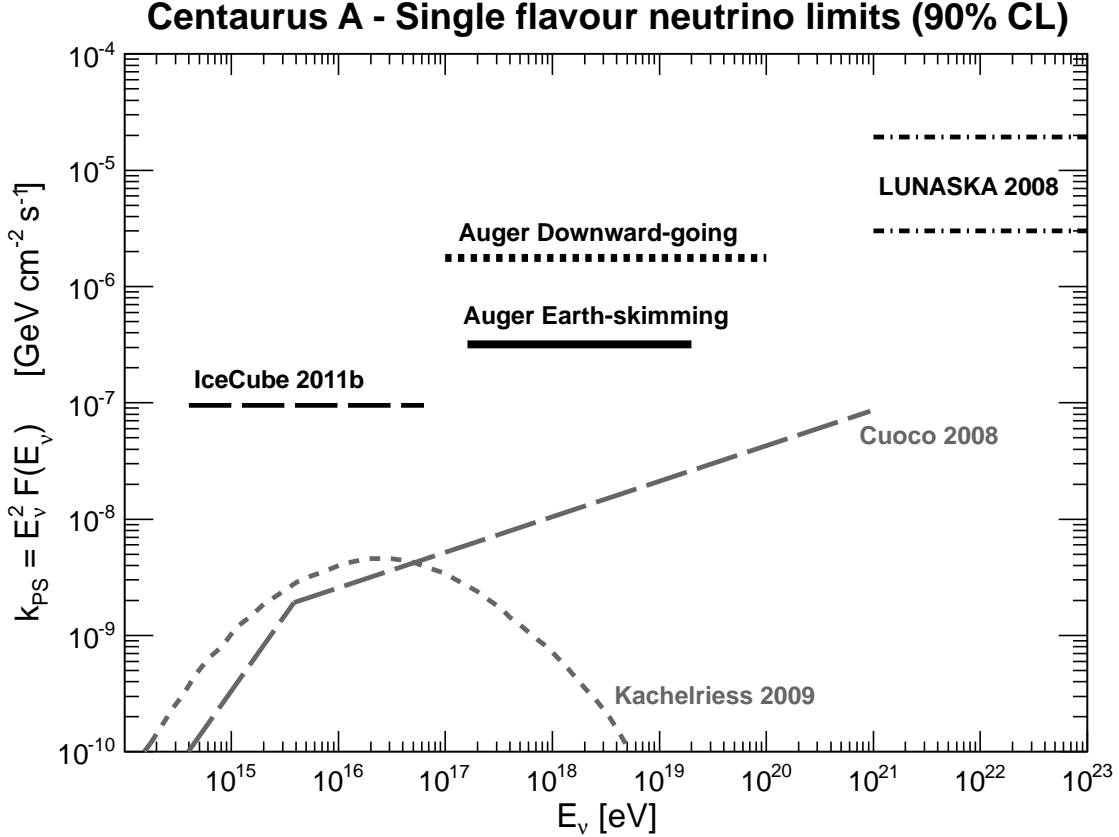}
\caption{Upper limits at 90$\%$ C.L.  on a single flavour $E^{-2}_\nu$ flux from the
active galaxy Centaurus A from the Earth-skimming and downward-going neutrino analyses,
together with bounds from the IceCube Neutrino Observatory~\cite{Abbasi11b}
and LUNASKA~\cite{LUNASKA11}.
The predictions for two models of UHE$\nu$ production -- in the jets~\cite{Cuoco08},
and close to the core of Centaurus A~\cite{Kachel09} -- are also shown.
\label{fig:CenA_limits}}
\end{center}
\end{figure}

Limits for the particular case of the active galaxy Centaurus A, a potential source of UHECRs, are
shown in Figure~\ref{fig:CenA_limits}, together with constraints from
other experiments. The predicted fluxes for two theoretical models of UHE$\nu$
production -- in the jets~\cite{Cuoco08} and close to the core
of Centaurus A~\cite{Kachel09} -- are also shown for comparison.
The expected number of events in our blind search samples for a flux like in~\cite{Cuoco08} is about 0.1 and 0.02  for the {\it ES} and {\it DG} selection respectively, the expected number for~\cite{Kachel09} being one order of magnitude smaller.

\section{ SUMMARY }
\label{sec:Conclusions} The sensitivity of the Pierre Auger
Observatory to point-like sources of neutrinos with ultra
high-energy has been obtained. Highly inclined extensive air showers
produced by the interaction of neutrinos of all flavours in the
atmosphere  and by the decay of tau leptons  originating from tau
neutrinos interactions in the Earth crust were searched for, and
discriminated from the background of standard UHECRs thanks to the
distinctive time structure of the signals measured by the Surface
Detector array. The search for neutrinos was performed over a broad
range of declination, north of $\sim -65^\circ$ and south of $\sim
55^\circ$, and for neutrino energies between $10^{17}$~eV and
$10^{20}$~eV.

No neutrino candidates were found in data collected through 2010 May
31, and  an improved upper limit on the diffuse flux of tau
neutrinos was correspondingly placed. Also, the exposure for
neutrino fluxes from point-like sources was evaluated as a function
of source declination. Upper limits at 90\% C.L. for neutrino fluxes
from point-like sources were established, which are currently the
most stringent at energies around and above 1 EeV in a large
fraction of the sky spanning more than 100$^\circ$ in declination.

\section{ Acknowledgements}
\label{sec:Acknow}
The successful installation, commissioning, and operation of the Pierre Auger Observatory
would not have been possible without the strong commitment and effort
from the technical and administrative staff in Malarg\"ue.

We are very grateful to the following agencies and organizations for financial support:
Comisi\'on Nacional de Energ\'ia At\'omica,
Fundaci\'on Antorchas,
Gobierno De La Provincia de Mendoza,
Municipalidad de Malarg\"ue,
NDM Holdings and Valle Las Le\~nas, in gratitude for their continuing
cooperation over land access, Argentina;
the Australian Research Council;
Conselho Nacional de Desenvolvimento Cient\'ifico e Tecnol\'ogico (CNPq),
Financiadora de Estudos e Projetos (FINEP),
Funda\c{c}\~ao de Amparo \`a Pesquisa do Estado de Rio de Janeiro (FAPERJ),
Funda\c{c}\~ao de Amparo \`a Pesquisa do Estado de S\~ao Paulo (FAPESP),
Minist\'erio de Ci\^{e}ncia e Tecnologia (MCT), Brazil;
AVCR AV0Z10100502 and AV0Z10100522, GAAV KJB100100904, MSMT-CR LA08016,
LG11044, MEB111003, MSM0021620859, LA08015 and TACR TA01010517, Czech Republic;
Centre de Calcul IN2P3/CNRS,
Centre National de la Recherche Scientifique (CNRS),
Conseil R\'egional Ile-de-France,
D\'epartement  Physique Nucl\'eaire et Corpusculaire (PNC-IN2P3/CNRS),
D\'epartement Sciences de l'Univers (SDU-INSU/CNRS), France;
Bundesministerium f\"ur Bildung und Forschung (BMBF),
Deutsche Forschungsgemeinschaft (DFG),
Finanzministerium Baden-W\"urttemberg,
Helmholtz-Gemeinschaft Deutscher Forschungszentren (HGF),
Ministerium f\"ur Wissenschaft und Forschung, Nordrhein-Westfalen,
Ministerium f\"ur Wissenschaft, Forschung und Kunst, Baden-W\"urttemberg, Germany;
Istituto Nazionale di Fisica Nucleare (INFN),
Ministero dell'Istruzione, dell'Universit\`a e della Ricerca (MIUR), Italy;
Consejo Nacional de Ciencia y Tecnolog\'ia (CONACYT), Mexico;
Ministerie van Onderwijs, Cultuur en Wetenschap,
Nederlandse Organisatie voor Wetenschappelijk Onderzoek (NWO),
Stichting voor Fundamenteel Onderzoek der Materie (FOM), Netherlands;
Ministry of Science and Higher Education,
Grant Nos. N N202 200239 and N N202 207238, Poland;
Funda\c{c}\~ao para a Ci\^{e}ncia e a Tecnologia, Portugal;
Ministry for Higher Education, Science, and Technology,
Slovenian Research Agency, Slovenia;
Comunidad de Madrid,
Consejer\'ia de Educaci\'on de la Comunidad de Castilla La Mancha,
FEDER funds,
Ministerio de Ciencia e Innovaci\'on and Consolider-Ingenio 2010 (CPAN),
Xunta de Galicia, Spain;
Science and Technology Facilities Council, United Kingdom;
Department of Energy, Contract Nos. DE-AC02-07CH11359, DE-FR02-04ER41300,
National Science Foundation, Grant No. 0450696,
The Grainger Foundation USA;
NAFOSTED, Vietnam;
ALFA-EC / HELEN
and UNESCO.

\clearpage

\clearpage

\clearpage

\clearpage

\clearpage

\clearpage


\end{document}